\documentclass[twocolumn,aps,showpacs,pre,superscriptaddress]{revtex4}
\usepackage{hyperref}
\usepackage[dvips]{graphicx}
\usepackage{multirow}
\usepackage{amsmath,amssymb,latexsym,MnSymbol}

\begin{document}

\title{Anomalous Diffusion and Long-range Correlations in the Score Evolution of the Game of Cricket}

\author{Haroldo V. Ribeiro}\email{hvr@dfi.uem.br}
\affiliation{Departamento de F\'isica and National Institute of Science and Technology for
Complex Systems, Universidade Estadual de Maring\'a, Maring\'a, PR 87020, Brazil}
\affiliation{Department of Chemical and Biological Engineering, Northwestern University, Evanston, IL 60208, USA}
\author{Satyam Mukherjee}
\affiliation{Department of Chemical and Biological Engineering, Northwestern University, Evanston, IL 60208, USA}
\author{Xiao Han T. Zeng}
\affiliation{Department of Chemical and Biological Engineering, Northwestern University, Evanston, IL 60208, USA}

\date{\today}

\begin{abstract}
We investigate the time evolution of the scores of the second most popular sport
in world: the game of cricket. By analyzing the scores event-by-event of more than
two thousand matches, we point out that the score dynamics is an anomalous diffusive process. 
Our analysis reveals that the variance of the process is described by a power-law 
dependence with a super-diffusive exponent, that the scores are statistically self-similar
following a universal Gaussian distribution, and that there are long-range correlations in the
score evolution. We employ a generalized Langevin equation with a power-law correlated noise
that describe all the empirical findings very well. These observations suggest that
competition among agents may be a mechanism leading to anomalous diffusion and long-range
correlation.
\end{abstract}

\pacs{02.50.-r,05.45.Tp,89.20.-a}

\maketitle

Diffusive motion is ubiquitous in nature. It can represent how
a drop of ink spreads in water, how living organisms 
such as fishes~\cite{Skalski} or bacteria~\cite{Ishikawa} move,
how information
travels over complex networks~\cite{Iribarren}, and many other phenomena.
One of the most common fingerprints of usual diffusion is the way that
particles or objects in question spread. The spreading can
be measured as the variance of the positions of the particles 
after a certain period of time. For usual diffusion, the variance 
grows linearly in time. There are two hypothesis underlying this behavior.
The first one is the absence of memory along the particle trajectory, that is, 
the actual position of the particle can be approximated by a function of its 
immediately previous position (Markovian hypothesis). The second one is the 
existence of a characteristic scale for the position increments. 
When these two assumptions hold, we can show that distribution of the 
positions will approach a Gaussian profile (Central Limit Theorem).

Naturally, there are situations in nature that do not fit these hypothesis
and, consequently, deviations from the usual behavior appear. When this happens, 
researchers usually report on anomalous diffusion. A well understood
case is when there is no characteristic length for the particle jumps. In this
case, the variance is infinity and the distribution of the positions follows
a L\'evy distribution. Examples of L\'evy processes include the animals movement 
during foraging~\cite{Humphries}, diffusion of ultracold atoms~\cite{Sagi} 
and systems out of thermal equilibrium~\cite{Bardou}. 
The situation is more complex when the diffusive process presents memory. 
We have many different manners of correlating the particle positions.
Depending on this choice, diffusive properties such as the dependence of
the variance in time can drastically change. In this context,
a typical behavior for the variance is a power-law dependence with an exponent
$\alpha$, where $\alpha<1$ corresponds to sub-diffusion and $\alpha>1$ to
super-diffusion.

Several approaches have been proposed to investigate anomalous diffusion in general. 
Fractional diffusion equations~\cite{Metzler}, Fokker-Planck equations~\cite{Trigger},
and Langevin equations~\cite{Lim} are just a few examples of frameworks used to describe 
this phenomenon. However, there is a lack of empirical studies aiming to verify
situations where these models can be applied and the possible mechanisms 
that lead to anomalous diffusion. There are a few exceptions, such as the work of 
Weber, Spakowitz and Theriot~\cite{Weber} where they showed that the motion of
chromosomal loci of two bacterial species is sub-diffusive and anti-correlated,
as well as the work of Lenz \textit{et al.}~\cite{Lenz} where they investigated the
role of predation in the motion of bumblebees during foraging.

In this work, we show that the evolution of the scores in the game
of cricket can be understood as a diffusive process with scale-invariance
properties, anomalous diffusion, and long-range correlations. All these
findings are well described by a generalized Langevin equation with
a power-law correlated noise. The results presented here suggest that
competition among agents may be a mechanism leading to correlation and anomalous diffusion
correlation. In the following, we present our dataset of scores of 
cricket matches, a diffusive interpretation for the evolution of these 
scores, a generalized Langevin equation
for modeling the empirical findings, and finally, some concluding remarks. 

The game of cricket is the second most popular sport in the world after
soccer. It is a ``bat-and-ball'' game (similar to baseball) played between two
teams of 11 players. There are three types of the game that differ in length.
The ``Twenty20'' (T20) cricket is the shortest one lasting approximately $3$ hours,
the ``One Day International'' (ODI) cricket lasts almost $8$ hours, and the ``Test'' 
cricket is the longest one taking up to five days to finish. The game involves one team 
batting (their innings) and scoring as many points (runs) as possible and setting up a 
target for the opponent team. The opponent team comes in to bat and tries to 
exceed the target. A team's innings is terminated whenever it exceeds the 
quota of \textit{overs} (six consecutive balls bowled in succession) or when the team
lost 10 \textit{wickets} (wooden stumps used as a target for the bowling).
The maximum limit is $20$ overs for T20 cricket, $50$ overs for ODI cricket, and 
$200$ overs for Test cricket.

\begin{figure}[!t]
\centering
\includegraphics[scale=0.43]{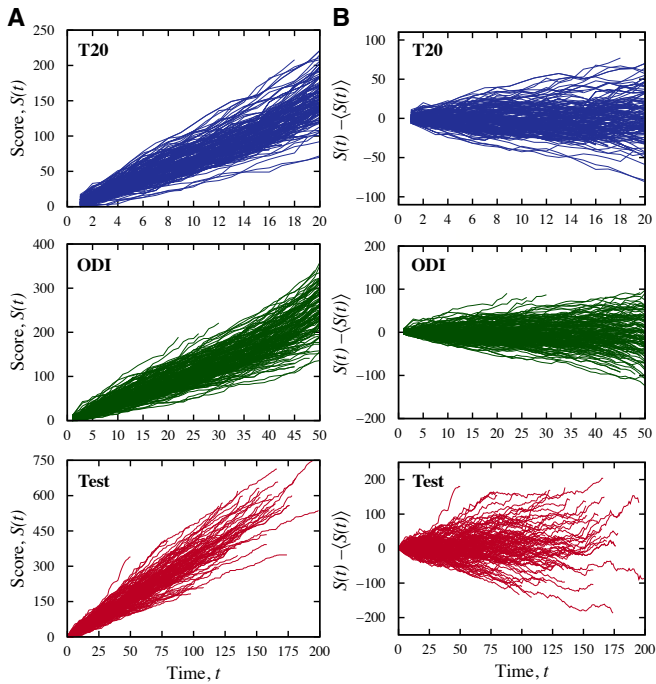}
\caption{Evolution of the scores $S(t)$ for the different three types of cricket.
To make the notation easier, we have denoted the event-by-event evolution as time evolution.
The main difference between these types is the maximum length of game. 
The maximum length is 20 time steps for T20, 50 for ODI, and 200 for test. 
In panel (A) we plot the 
evolution of the scores of a hundred games selected at random from our database, and in panel
(B) we show these evolution after removing the mean tendency of increase, that is, $S(t)-\langle S(t)\rangle$.
}\label{fig1}
\end{figure}

Surprisingly, the record of a game of cricket (score cards) includes not only the game outcome,
but also the event-by-event evolution of the scores. We collect the information of 
scores per over for T20 ($2005-2011$), ODI ($2002-2011$), and Test cricket ($2002-2011$)  
from the \textit{cricinfo} website~\cite{Site}. Using these data, we create 2144 time series 
of scores where the time $t$ represents a completed \textit{over}. In Figure~\ref{fig1}A,
we show the temporal dependence of the scores $S(t)$ for one hundred games
from the three different types of cricket selected at random from our database.
We note the natural increasing tendency of the scores and also the erratic movement
around the mean tendency. For better visualization  of these
fluctuations, we plot in Fig.~\ref{fig1}B the scores after subtracting the 
mean tendency $\langle S(t) \rangle$ from $S(t)$.

\begin{figure}[!t]
\centering
\includegraphics[scale=0.43]{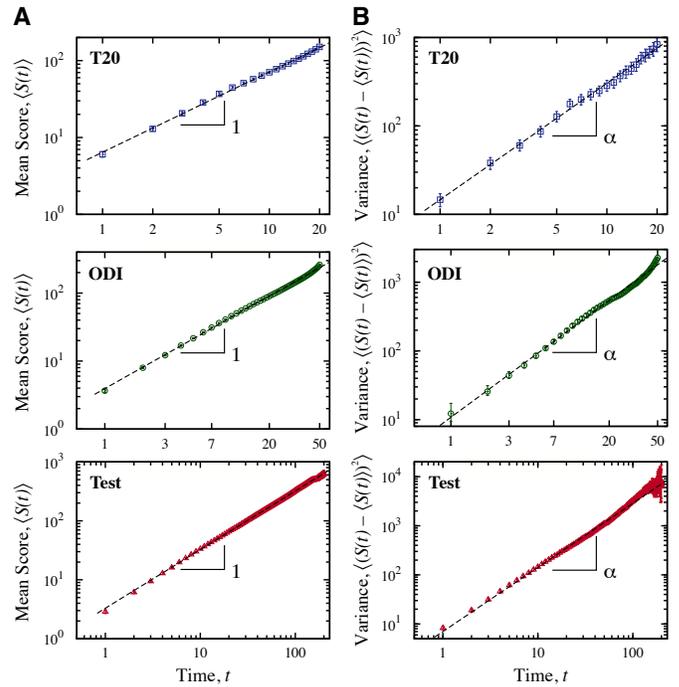}
\caption{The anomalous diffusion of the scores. (A) Mean value of the scores as a function 
of time, $\langle S(t)\rangle$, for the three types of the Cricket.
The dashed lines are linear fits to each dataset. We find the mean values to grow linearly 
in time. (B) The spreading of the score trajectories measured as the variance 
$\sigma^2(t)=\langle [S(t) - \langle S(t)\rangle]^2 \rangle$ versus
time. The dashed lines are power-law fits to the variance, where we find the 
exponents $\alpha=1.32\pm0.02$ for T20, $\alpha=1.31\pm0.02$ for ODI, and $\alpha=1.30\pm0.02$ 
for Test. 
Since $\alpha>1$, the diffusive process 
underlying the evolution of scores is super-diffusive. The error bars are $95\%$ confidence
intervals obtained via bootstrapping~\cite{Efron}.
}\label{fig2}
\end{figure}

We start by investigating how the mean value of the scores depends on time
(Fig.~\ref{fig2}A). These plots reveal that the mean
score $\langle S(t)\rangle$ grows linearly in time for the three types of 
the game. The only difference is in the rate of growth, which is $6.4\pm1.0$ for T20, 
$3.9\pm1.0$ for ODI, and $3.3\pm1.0$ for Test. The different values show that
the overall performance of teams are related to length of the game. In T20 cricket 
(which last $\sim3$ hours) and in ODI cricket (which last $\sim8$ hours), we have the
highest rates indicating that the players work hard for scoring 
as many points as they can, while for Test cricket the players may prefer to save efforts, 
since Test matches are quite long.
 
Next, we characterize the spreading process by evaluating
the variance of the scores as a function of time (Fig.~\ref{fig2}B). 
We show the variance $\sigma^2(t)=\langle [S(t) - \langle S(t)\rangle]^2 \rangle$ 
in a log-log plot, where we observe a non-linear increase of $\sigma^2(t)$.
By least square fitting a linear model to these log-log data, we find
a super-diffusive regime, that is, $\sigma^2(t)\propto t^\alpha$ with $\alpha\approx 1.3$ 
for the three types of cricket. This intriguing feature suggests that the 
competition within the game may drive the scores to spread faster than a regular 
Brownian motion. 

\begin{figure}[!t]
\centering
\includegraphics[scale=0.43]{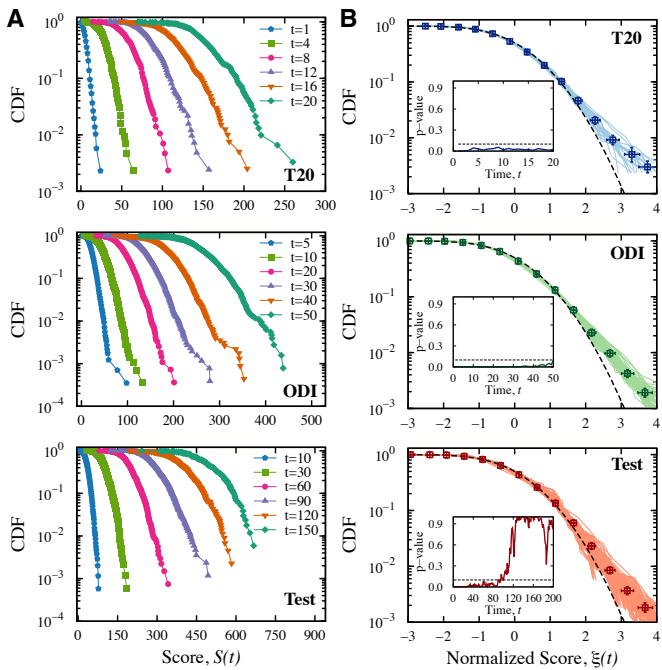}
\caption{Scale invariance of the scores. 
(A) Evolution of the cumulative distribution function (CDF) for the three types of 
cricket and for different values of time $t$. Note that the distributions shift towards 
positive values of the score and that the width of the CDFs increases. (B) Scale invariance 
of the scores. We evaluate the CDF using the normalized scores 
$\xi(t)=\frac{S(t)-\langle S(t)\rangle}{\sigma(t)}$ where $\langle S(t)\rangle$ is the 
mean value of the scores and $\sigma(t)$ is the square root of the variance of the scores. 
Note that the good collapse of the distributions indicates that the scores present
scaling properties, \textit{i.e.}, after normalization they follow the same universal 
distribution. In these plots, the continuous lines are the CDFs for each value of $t$ 
and the symbols are the averaged values of these CDFs. The error bars are $95\%$ confidence
intervals obtained via bootstrapping~\cite{Efron}. We note further that these distributions 
are very close to a normalized Gaussian distribution (dashed lines). The insets show
the p-values for the Pearson chi square test~\cite{Greenwood} as a function of time. The
dashed line is the threshold $0.1$ for rejecting the Gaussian hypothesis. 
Note that the
normality is rejected for small values of $t$ because of the discrete values of $S(t)$
and also the asymmetric initial condition of the diffusive process. After enough time,
($t\sim 90$), we can not reject the Gaussian hypothesis in the Test cricket.
}\label{fig3}
\end{figure}

Another interesting question is whether the distribution of the scores is self-similar
and whether these distributions follow a particular functional form. To answer this question,
we calculate the cumulative distribution functions of the scores for each time step. 
Figure~\ref{fig3}A shows these distributions for several values of $t$ and for the three
types of cricket. We note the shift of the distributions towards positive values 
and the increase in the distribution width. Moreover, these semi-log plots indicates
that the distributions are close to normal distributions. 

To check the normality and self-similarity, we evaluate the distribution of the normalized 
scores $\xi(t)=\frac{S(t)-\langle S(t)\rangle}{\sigma(t)}$, where $\langle S(t)\rangle$ is
the mean value of the score and $\sigma(t)$ is the standard-deviation. As shown in 
Fig.~\ref{fig3}B, the distributions exhibit a good collapse and a profile that is very
close to Gaussian distribution. These results are also supported by the insets of Fig.~\ref{fig3}B,
where we plot the p-values for the Pearson chi square test as a function of time.
We note that the normality is rejected for small values of $t\lesssim90$ due to the 
discrete nature of $S(t)$ and also the asymmetry in the score system. After enough time 
($t\sim90$), the p-values are larger than $0.1$ and we can not reject the Gaussian hypothesis
in the Test cricket.

\begin{figure}[!t]
\centering
\includegraphics[scale=0.52]{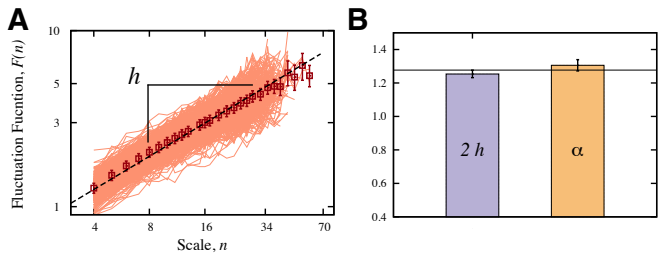}
\caption{Long-range correlations in the scores. (A) Detrended fluctuation 
analysis of score increments $\Delta S(t) = S(t+1)-S(t)$. We show the fluctuation 
functions $F(n)$ versus the scale $n$ (continuous lines) for all games 
of the Test cricket that are longer than $120$ units of time (431 games). Note that $F(n)$ 
follows a power-law, where the exponent $h$ is the Hurst exponent. We estimate the mean value 
of Hurst exponent to be $h=0.63\pm0.01$, and the dashed line is a power-law with this exponent.
We find the average value of Pearson linear correlation coefficient equal to $0.89\pm0.02$, 
which enhances the quality of the power-law relationships. The symbols represent the average
values of the fluctuation functions and the error bars are standard errors of the means.
(B) Comparison of the model prediction, that is, $\alpha=2 h$ for the Test cricket. The left
bar shows the empirical value of $2 h$ and the right bar shows the value of $\alpha$. The error
bars are $95\%$ confidence intervals and the horizontal line is the upper limit of 
the confidence interval for $2 h$. We note the existence of overlapping in the confidence intervals,
indicating that the relation $\alpha=2 h$ holds. 
}\label{fig4}
\end{figure}

We now focus on correlation analysis to answer whether the scores evolution
is a Markovian process. To investigate this hypothesis, we select all games from the
Test cricket that are longer than 120 time steps, totaling 431 games. 
For this subset, we calculate the time series of the scores increments $\Delta S(t)=S(t+1)-S(t)$. 
Next, we employ detrended fluctuation analysis (DFA) to obtain the Hurst exponent $h$.
DFA~\cite{Peng,Kantelhardt} consists of four steps: $i)$ We first define the profile
$Y(i)=\sum_{k=1}^{i} \Delta S(t) - \langle \Delta S(t) \rangle\,.$ $ii)$ Next, we cut $Y(i)$ 
into $N_n=N/n$ non-overlapping segments of size $n$, where $N$ is the length of the series.
$iii)$ For each segment, a local polynomial trend (here we have used a linear function) 
is calculated and subtracted from $Y(i)$, defining $Y_n(i)=Y(i)-p_\nu(i)$, 
where $p_\nu(i)$ represents the local trend in the $\nu$-th segment.
$iv)$ Finally, we evaluate the root-mean-square fluctuation function
$F(n)=[\frac{1}{N_n}\sum_{\nu=1}^{N_n} \langle Y_n(i)^2\rangle_\nu]^{1/2}\,,$
where $\langle Y_n(i)^2\rangle_\nu$ is mean square value of $Y_n(i)$ over the 
data in the $\nu$-th segment.
For self-similar time series, the fluctuation function $F(n)$ displays
a power-law dependence on the time scale $n$, that is, $F(n)\sim n^{h}$, where $h$
is the Hurst exponent. Intriguingly, we find that the Hurst exponent does not depend on game
and that it has a mean value equal to $h=0.63\pm0.01$ (Fig.~\ref{fig4}A). This result shows that there is long-range
memory in the score evolution, and therefore it is a non-Markovian process. Moreover,
the value of $h>0.5$ indicates the existence of a persistence behavior in the scores 
increments, that is, positive values are followed by positive values and negative
values are followed by negative values much more frequently than by chance.

All the previous empirical findings claim for model. To address this question, we consider
the following generalized Langevin equation for describing the score evolution of the Test cricket,
\[
\frac{d^2 S(t)}{dt^2}+\int_0^t \lambda(t-\tau)\frac{d S(\tau)}{d\tau} d\tau+K=\xi(t)\,.
\]
Here, $\lambda(t-\tau)$ is the retarded effect of the frictional force, $K$ is a drift constant, 
and $\xi(t)$ represents a Gaussian stochastic force. Because we know that long-range correlations are
present in our system, we consider that $\xi(t)$ is also power-law correlated, that
is, $\langle \xi(0) \xi(t)\rangle \sim t^{-\alpha}$. We also assume 
$\lambda(t)\propto \langle \xi(0) \xi(t)\rangle$ in order to satisfy the 
fluctuation-dissipation theorem~\cite{Kubo}. This equation was
presented in Refs.~\cite{Lim,Wang,Wang2} for $K=0$ and it can be solved by using
Laplace transform. Indeed, after some calculations, we can show that the mean score is 
linear in time $\langle S(t) \rangle \sim t$, that the variance obeys a power-law relationship 
$\langle [S(t) - \langle S(t)\rangle]^2 \rangle\sim t^{-\alpha}$, and the distribution of the
scores is Gaussian. Remarkably, these are exactly the same features that our empirical 
data present (see Figs.~\ref{fig2} and \ref{fig3}).

Furthermore, we calculate the auto-correlation function of the score ``velocity''
$\langle V(0)V(t) \rangle \sim t^{\alpha-2}$, where $V(t)=dS(t)/dt$ and $\alpha\neq1$.
Note that this derivative corresponds to the score increments $\Delta S(t)=S(t+1)-S(t)$
in the discrete case. Thus, the Langevin equation also predicts the existence of long-range
memory in the score increments. We can check this prediction by observing 
that the power-law exponent of auto-correlation function is related to the Hurst 
exponent~\cite{Kantelhardt}, which consequently leads to a relationship between
the Hurst and the diffusive exponent $\alpha=2h$. Figure~\ref{fig4}B shows a bar plot
that compares the value of $2h$ (left bar) with the value of $\alpha$ (right bar) for the 
Test cricket. We note that these values are close to each other and that there exists overlapping 
between the confidence intervals. Therefore, the relationship $\alpha=2h$ applies.

In summary, we have studied the score evolution of cricket games as a diffusive process.
Our analysis reveals that the mean score grows linearly in time, while the variance of the
scores has a power-law dependence in time with a super-diffusive exponent. We show 
that the scores are statistically self-similar and follow a universal distribution approximated by a Gaussian.
By using DFA, we point out that this diffusive process is non-Markovian since the
scores increments are long-range correlated. 
It is worth to note that the persistent 
long-range memory present in the diffusive process can be related the ``hot hand'' 
phenomenon in sports. Since the seminal work of Gilovich et al.~\cite{Gilovich} there
has been a historical debate on whether ``success breeds success'' or ``failure breeds failure'' 
in the scoring process of many sports~\cite{Yaari,Yaari2}. Here, the long-range 
persistent behavior in the score evolution not only indicates the existence of 
this phenomenon in cricket, but also suggests that this phenomenon can act over 
a very long temporal scale. Because of the long-range memory, we proposed to model the
empirical findings using a generalized Langevin equation driven by a power-law correlated 
stochastic force. The correlation in the noise term induces the faster-than-regular
spreading of the diffusive process and also gives rise to correlations in the score 
increments. The results of this model show that there is a simple relation between 
the diffusive exponent $\alpha$ and the Hurst exponent $h$, which we have verified
to hold in the empirical data. We are optimistic that the discussion presented may be
applied to other sports, where new analysis can reveal more complex diffusive patterns
to be compared with the increasing number of theoretical results on anomalous diffusion.

\acknowledgements
We thank the \textit{cricinfo} website for making publicly available the data set used here.
Fruitful discussions with members of the Amaral Lab are also gratefully acknowledged.
HVR is grateful to CAPES for financial support under the process No~5678-11-0.

\end{document}